\documentclass[journal=nalefd,manuscript=article]{achemso}
\usepackage{graphicx}
\usepackage{amsmath,amssymb}
\usepackage{epstopdf}
\usepackage{booktabs}

\author{Jun-Jie Zhang}
\affiliation{School of Physics, Southeast University, Nanjing 211189, China}
\alsoaffiliation{Department of Materials Science and NanoEngineering, Rice University, Houston, Texas 77005, United States}
\author{Linfang Lin}
\author{Yang Zhang}
\affiliation{School of Physics, Southeast University, Nanjing 211189, China}
\author{Menghao Wu}
\affiliation{School of Physics and Wuhan National High Magnetic Field Center, Huazhong University of Science and Technology, Wuhan 430074, China}
\author{Boris I. Yakobson}
\email{biy@rice.edu}
\affiliation{Department of Materials Science and NanoEngineering, Rice University, Houston, Texas 77005, United States}
\author{Shuai Dong}
\email{sdong@seu.edu.cn}
\affiliation{School of Physics, Southeast University, Nanjing 211189, China}

\title{Type-II multiferroic Hf$_{2}$VC$_{2}$F$_{2}$ MXene monolayer with high transition temperature}


\begin{document}
\begin{abstract}
Achieving multiferroic two-dimensional (2D) materials should enable numerous functionalities in nanoscale devices. Until now, however, predicted 2D multiferroics are very few and with coexisting yet only loosely coupled (type-I) ferroelectricity and magnetism. Here, a type-II multiferroic MXene Hf$_{2}$VC$_{2}$F$_{2}$ monolayer is identified, where ferroelectricity originates directly from its magnetism. The noncollinear Y-type spin order generates a polarization perpendicular to the spin helical plane. Remarkably, the multiferroic transition is estimated to occur above room temperature. Our investigation should open the door to a new branch of 2D materials for pursuit of intrinsically strong magnetoelectricity.
\end{abstract}

Two-dimensional (2D) materials have attracted attention since the experimental discovery of graphene \cite{novoselov2004electric}. Divers properties and functions are discovered in plentiful 2D materials, going beyond the original appeal as new semiconductors. More and more important physical properties existing in three-dimensional (3D) crystals have also been found to appear in the 2D form. For example, 2D superconductivity \cite{ge2015superconductivity,zhang2016strain,zhang2016blue,zhang2017superconductivity,lei2017predicting}, 2D ferromagnetism \cite{huang2017layer,dong2017rational,yang2016tunable}, as well as 2D ferroelectricity \cite{wu2013hydroxyl,kan2013high,shirodkar2014emergence,di2015emergence,ding2017prediction,wu2016ferroelectricity,wu2016intrinsic,chandrasekaran2017ferroelectricity,fei2016ferroelectricity,li2017binary,chang2016discovery,liu2016room}, have recently been confirmed in experiments or predicted via calculations.

Ferromagnetism and ferroelectricity, with switchable ferro-vectors, play crucial roles in various devices. Thus, their existence in 2D materials would be very attractive. On one hand, since the first prediction of 2D ferroelectric (FE) hydroxylized graphene in 2013 \cite{wu2013hydroxyl}, more 2D materials, e.g. 1T-MoS$_{2}$, In$_2$Se$_3$, 2D materials functionalized with polar groups, etc., have been predicted to be FE \cite{kan2013high,shirodkar2014emergence,di2015emergence,ding2017prediction,wu2016ferroelectricity,wu2016intrinsic,chandrasekaran2017ferroelectricity,fei2016ferroelectricity,li2017binary}. Experimentally, in-plane FE polarization ($P$) was observed and manipulated in atom-thick SnTe \cite{chang2016discovery}, while out-of-plane FE $P$ was found in few layes CuInP$_{2}$S$_{6}$ \cite{liu2016room}. The involved mechanisms are either polar phonon modes or polar functional groups. On the other hand, many 2D ferromagnets have been predicted \cite{dong2017rational,yang2016tunable}, and recently CrI$_3$ monolayer is experimentally confirmed \cite{huang2017layer}.

The coexisting magnetism and polarization lead to the multiferroicity \cite{cheong2007multiferroics,dong2015multiferroic}. The coupling between these two ferro-vectors allows the control of charge via magnetic field or the control of spin via electric field. In fact, a few 2D multiferroics were also recently predicted \cite{tu2017two,yang2017chemically,wu2016intrinsic,li2017binary}, in which the origins of polarization and magnetism are independent of each other (i.e. they are all type-I multiferroics \cite{Khomskii:Phy}). Thus their magnetoelectric coupling is indirect and weak.

To pursue the intrinsically strong magnetoelectricity, a possible route is to design 2D type-II multiferroics (i.e. magnetic ferroelectrics \cite{Khomskii:Phy}), in which the FE $P$ is directly generated and thus fully controlled by magnetic order \cite{cheong2007multiferroics,dong2015multiferroic}. Although the type-II multiferroics have been extensively studied, these materials have not gone into the zone of 2D materials. Even some type-II multiferroics own layered structures, e.g. CuFeO$_{2}$ \cite{kimura2006inversion}, Sr$_3$NiTa$_2$O$_9$ \cite{liu2016two}, and Ba$_3$MnNb$_2$O$_9$ \cite{lee2014magnetic}, the inter-layer couplings are via ionic bonds, difficult to be exfoliated.

In this work, monolayer carbides and carbonitrides, i.e. MXenes ($M_{n+1}X_{n}T_{x}$, $M$: early transition metal; $X$: carbon or nitrogen), are predicted as 2D type-II multiferroics. As a new branch of 2D materials, MXenes have been experimentally produced by selectively
etching the A-layers from their 3D parent compounds MAX \cite{naguib2011two,naguib2012two,naguib2013new}. The surface can be easily covered by functional groups (e.g. $T$=F, O, or/and OH), resulting in diversiform chemical and physical properties \cite{khazaei2013novel,zhang2017superconductivity}. Recently, an ordered double transition metal MAX with Cr-Al/Mo-Al bonding were synthesized, e.g. Cr$_{2}$TiAlC$_{2}$ \cite{liu2014cr2} and Mo$_{2}$TiAlC$_{2}$ \cite{anasori2015mo2tialc2}, in which a Ti-layer is sandwiched between two outer Cr/Mo carbide layers in the $M_{3}AX_{2}$ structure. Then ordered double transition metal carbides $M'_{2}MX_{2}$ and $M'_{2}M_{2}X_{3}$, e.g. Mo$_{2}$TiC$_{2}T_{x}$, Mo$_{2}$Ti$_{2}$C$_{2}T_{x}$ and Cr$_{2}$TiC$_{2}T_{x}$, were successfully realized by etching the Al-layers \cite{anasori2015two}. In addition, considering the transition metals involved, many MXenes should be intrinsically magnetic (at least from the theoretical viewpoint), and their magnetism depends on $M$ ($M'$) and $T$ \cite{dong2017rational,yang2016tunable}. For instance, Cr$_{2}$TiC$_{2}$F$_{2}$ and Cr$_{2}$TiC$_{2}$(OH)$_{2}$ are predicted to be antiferromagnetic (AFM), whereas Cr$_{2}$VC$_{2}$(OH)$_{2}$, Cr$_{2}$VC$_{2}$F$_{2}$, and Cr$_{2}$VC$_{2}$O$_{2}$ are ferromagnetic (FM) \cite{yang2016tunable}. Herein, derived from experimental Hf$_3$C$_2T_x$ monolayer \cite{zhou2017synthesis}, the ordered double transition metal carbides Hf$_{2}M$C$_{2}T_2$ monolayers (possibly realized via 3D parent Hf$_{2}M$AlC$_{2}$) \cite{Supp} are considered to be a 2D type-II multiferroics. Although both $M$ and $M'$ can be magnetic ions, here only the middle $M$ layer is considered to be magnetic.

\textit{Candidate 2D MXene.} According to the knowledge of type-II multiferroicity, some special frustrated magnetic orders, like noncolliear spiral magnetism or $\uparrow\uparrow\downarrow\downarrow$-type AFM order, may break the space inversion symmetry and thus lead to FE $P$ \cite{cheong2007multiferroics,dong2015multiferroic}. The in-plane geometry of $M$ ions is triangular, which is inherently frustrated if the nearest-neighbor (NN) exchange is AFM \cite{ratcliff2016magnetic,lin2016hexagonal,kimura2006inversion,liu2016two}. Thus, to find MXenes with NN AFM interaction is the first step. According to the Goodenough-Kanamori rule \cite{goodenough1958interpretation,kanamori1959superexchange}, the ions with half-filled $d$ shell usually lead to strong AFM exchanges. Besides, the half-filled Hubbard bands can lead to insulating, as required for ferroelectricity.

\begin{figure}
\centering
\includegraphics[width=0.6\textwidth]{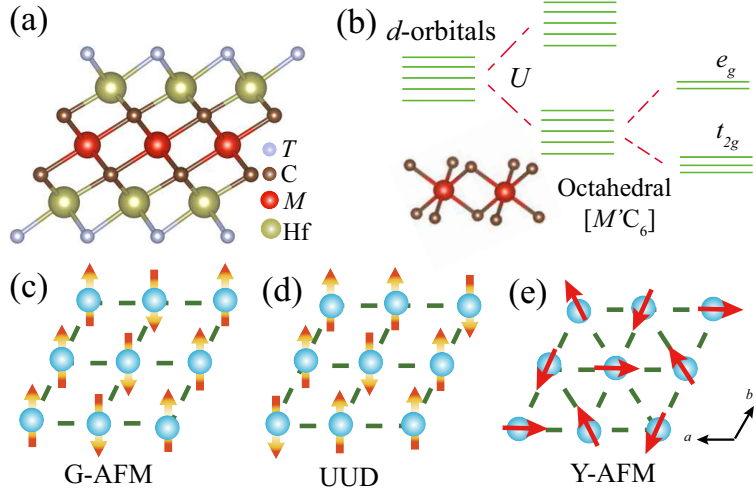}
\caption{(a) Side views of the AA CG for Hf$_{2}M$C$_{2}T_{2}$. (b) Sketch of the energy splitting of $3d$ orbitals for $M$. (c-e) Sketch of possible AFM and ferrimagnetic spin orders in the 2D triangular lattice.}
\label{Fig1}
\end{figure}

In $M'_{2}M$C$_{2}T_{2}$ MXene monolayer, each $M$C$_6$ forms an octahedron [\ref{Fig1}(a)]. The crystalline field of octahedron splits $M$'s $d$ orbitals into the low-lying $t_{\rm 2g}$ triplets and the higher-energy $e_{\rm g}$ doublets [\ref{Fig1}(b)]. To pursuit the half-filled Hubbard bands, high spin V$^{2+}$ ($3d^3$), Nb$^{2+}$ ($4d^3$), Mn$^{4+}$ ($3d^3$), Mn$^{2+}$ ($3d^5$), and Fe$^{3+}$($3d^5$) are possible candidates to play as $M$.

To satisfy aforemention conditions, in the following, Hf$_{2}$VC$_{2}$F$_{2}$, Hf$_{2}$NbC$_{2}$F$_{2}$, Hf$_{2}$MnC$_{2}$F$_{2}$, and Hf$_{2}$MnC$_{2}$O$_{2}$ will be calculated using density functional theory (DFT). Since Fe$^{3+}$ ($3d^5$) can not be obtained if symmetric $T$ layers are used, it is not considered here. In addition, $T$=OH will be not calculated, considering its equal valence (and thus similar physical effects) to F.

\textit{DFT results.}
DFT calculations are performed to verify the multiferroicity of these MXene monolayers. Our calculations find that Hf$_{2}$VC$_{2}$F$_{2}$ is the most possible 2D type-II multiferroic material, while others are unlike due to various reasons (see Supplemental Materials for more details of DFT methods and results \cite{Supp}). The possibility of synthesis for Hf$_{2}$VC$_{2}$F$_{2}$ is also explored, which maybe realized via 3D parent Hf$_{2}$VAlC$_{2}$. Based on experimentally produced V$_{3}$AlC$_{2}$ \cite{naguib2013new} and Hf$_{3}$AlC$_{2}$ \cite{lapauw2016synthesis}, the hybrid energy for Hf$_{2}$VAlC$_{2}$ is about $-90$ meV/cell, implying such mix is more favorable. In addition, the possibility of MAX phase to MXenes for Hf$_{2}$VAlC$_{2}$ are also verified by the crystal orbital Hamilton population (COHP) and exfoliation energies calculations \cite{Supp}. Based on the results of COHP, the obtained bond strength between Hf and Al is very weaker than other bond between Hf/V and C in Hf$_{2}$VAlC$_{2}$, which is similar to that in Hf$_{3}$AlC$_{2}$ and V$_{3}$AlC$_{2}$, indicating the iconicity/metallicity characteristic between them. To further examine the progress of exfoliation, exfoliation energies are calculated as E$_{exfoliation} = -[E_{tot}(MAX phase)-2E_{tot}(MXene)-E_{tot}(Al)]/(4S)$ \cite{khazaei2014effect}, where E$_{tot}$(MAX phase), E$_{tot}$(MXene) and E$_{tot}$(Al) stand for the total energies of bulk MAX phase, 2D MXene, and most stable bulk Al structure (Fm-3m), respectively. $S=\sqrt{3}a^2/2$ is the surface area and a is the lattice parameter of the MAX phase. Due to V$_{3}$AlC$_{2}$ was experimentally exfoliated into 2D Mxenes \cite{naguib2013new},  Hf$_{2}$VAlC$_{2}$ which ows lower exfoliation energies have a better possbility to be exfoliated into MXenes. In summary, we conclude Hf$_{2}$VAlC$_{2}$ is a good candidate of 3D parent phase for the successful exfoliation into 2D Hf$_{2}$VC$_{2}$ MXenes. More details can be found in Supplemental Materials \cite{Supp}.

First, various configurations (CG's) for Hf$_{2}$VC$_{2}$F$_{2}$ are verified. Based on energy comparison and dynamic stability, the AA CG is confirmed to be the most favorable one \cite{Supp}, where F ions stand just above/below the V's positions [\ref{Fig1}(a)]. Thus, our following investigation will focus on the AA CG only.

The nominal valences for Hf, V, C, and F are $+4$, $+2$, $-4$, and $-1$, respectively. Then for both C and F, the $2p$-orbitals are fully occupied, while for Hf the $5d$ orbitals are fully empty. In this sense, the magnetism can only come from V, whose $3d$ orbitals own three electrons, as confirmed in the DFT calculation.

\begin{figure}
\centering
\includegraphics[width=0.5\textwidth]{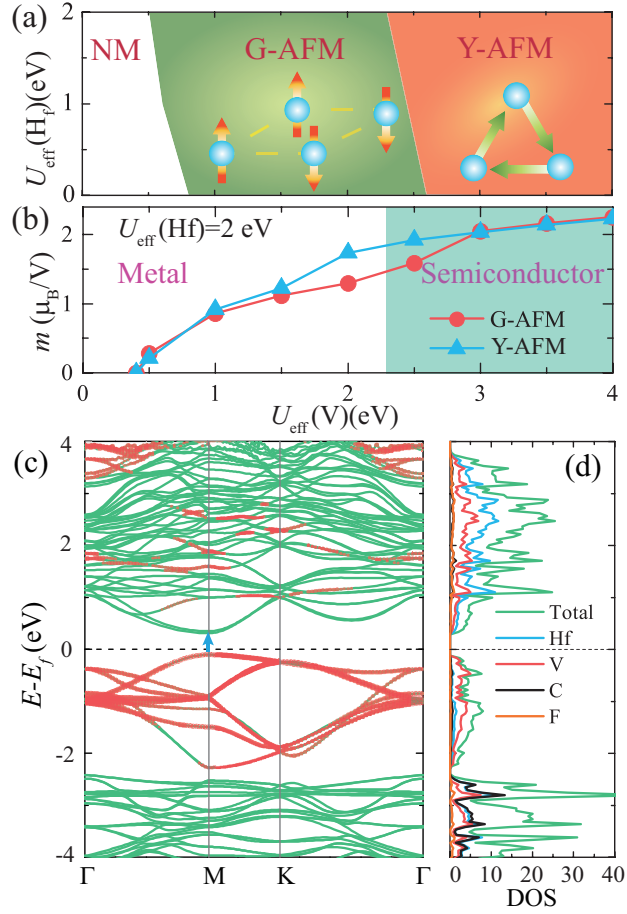}
\caption{DFT results of Hf$_{2}$VC$_{2}$F$_{2}$ as a function of $U_{\rm eff}$(Hf) and $U_{\rm eff}$(V). (a) The ground state phase diagrams. (b) Local magnetic moment of V for G- and Y-AFM calculated within the default Wigner-Seitz sphere. Metallic and insulating regions are distinguished by colors. (c) The electronic band structure for Y-AFM with SOC calculated at $U_{\rm eff}$(V)=$3$ eV and $U_{\rm eff}$(Hf)=$2$ eV. The fat bands (red circles) are contributed (more than $50\%$) by $3d$ orbitals of V ions. (d) The corresponding density of states (DOS) and atom-projected DOS (PDOS).}
\label{Fig2}
\end{figure}

The magnetic ground state of Hf$_{2}$VC$_{2}$F$_{2}$ is searched by comparing the energies of various possible magnetic orders, including the nonmagnetic (NM), collinear FM, UUD type ferrimagnetic (stands for the "up-up-down" ferrimagnetic spin order), stripe AFM (G-AFM), as well as the $120^\circ$ noncollinear AFM order (coined as Y-AFM here), as sketched in \ref{Fig1}(c-e). 

Considering the Hubbard-type correlations and spin-orbit coupling (SOC) for $3d$ and $5d$ orbitals, here a wide parameter space of $U_{\rm eff}$(V) and $U_{\rm eff}$(Hf) are scanned, as shown in ~\ref{Fig2}(a) and Supplemental Materials \cite{Supp}. As expected, the $U_{\rm eff}$(Hf) and SOC only have tiny effects on the magnetism due to Hf's empty $5d$ orbitals \cite{Supp}. In contrast, with increasing $U_{\rm eff}$(V), the magnetic ground-state undergoes two transitions, from NM to G-AFM first, then finally to Y-AFM. The local magnetic moment of V also depends on $U_{\rm eff}$(V), increasing from $0$ to more than $2$ $\mu_{\rm B}$/V (~\ref{Fig2}(b)). Accompanying the second magnetic transition, the metal-insulator transition also occurs when $U_{\rm eff}$(V)$>2$ eV.

\begin{table}
\centering
\caption{The calculated energies ($E$) of different magnetic structures using the HSE06 functional with SOC. The energy of FM is set as the reference. The corresponding local magnetic moments ($m$) are also listed.}
\begin{tabular*}{0.55\textwidth}{@{\extracolsep{\fill}}lccccc}
\hline
\hline
$~$       &NM      &FM      &G-AFM    &UUD  & Y-AFM  \\
\hline
$E$ (eV/V)   &$0.09$  &$0$    &$-0.12$  &$-0.14$  &$-0.19$ \\
$m$ ($\mu_{\rm B}$/V)      &$~$     &$1.18$  &$2.05$  &$1.99$ &$2.04$  \\
\hline
\hline
\end{tabular*}
\label{Table1}
\end{table}

Due to the lack of experimental result on Hf$_{2}$VC$_{2}$F$_{2}$ monolayer, the HSE06 functional with SOC, are adopted as the benchmark to provide an alternative description \cite{gou2011lattice}. As shown in ~\ref{Table1}, the HSE06 plus SOC calculation predicts that the Y-AFM is the ground state for Hf$_{2}$VC$_{2}$F$_{2}$ monolayer, and the corresponding magnetic moment is in good agreement with the result of $U_{\rm eff}$(V)$=3$ eV and $U_{\rm eff}$(Hf)$=2$ eV, implying this set of parameters is proper. In fact, the same $U_{\rm eff}$ parameters were also adopted in previous studies \cite{dong2017rational}, where only FM and G-AFM were considered. Here, four mostly-possible ones have been considered in DFT calculations. Furthermore, the following Monte Carlo simulation, with no bias of preset magnetic configurations, will be employed to verify the results of DFT. If there's more stable one, the Monte Carlo simulation should capture it.

The calculated electronic structure of Y-AFM with SOC are shown in ~\ref{Fig2}(c-d). It is clear that Hf$_{2}$VC$_{2}$F$_{2}$ monolayer is a direct-gap semiconductor and the corresponding band gap is about $0.4$ eV with default $U_{\rm eff}$'s. The HSE06 functional calculation leads to very similar electronic structure with a larger band gap ($0.9$ eV) \cite{Supp}. The projection of Bloch states to V's $d$-orbital is also displayed in ~\ref{Fig2}(c-d). As expected, there are nine occupied bands near the Fermi level mostly contributed by V's $3d$ orbitals. As expected, the $d^3$ configuration of V$^{2+}$ just occupies the $t_{\rm 2g}$ orbitals in the half-filling manner, while the $e_{\rm g}$ orbitals are above the Fermi level. According to PDOS, there is also moderate $p$-$d$ hybridization, which is a bridge for superexchange interaction.

\begin{figure}
\centering
\includegraphics[width=0.6\textwidth]{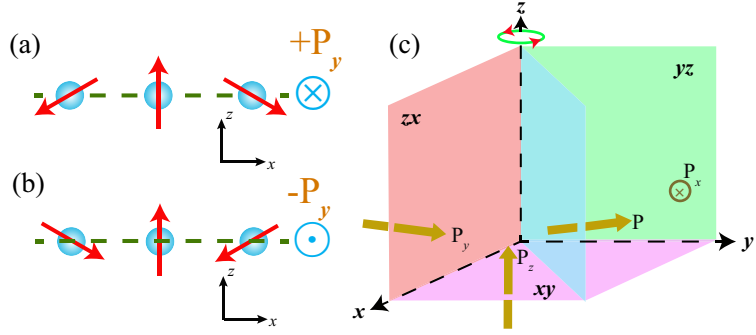}
\caption{Schematic of helical Y-AFM induced FE $\textbf{P}$. (a-b) Clockwise \textit{vs} counter-clockwise helicity. The corresponding $\textbf{P}$ is perpendicular to the Y-AFM spin plane. (c) The free rotation of spin plane along the $c$-axis ($z$-axis). The higher energy $ab$ ($xy$) plane is also shown. Here $x$-$y$-$z$ forms rectangular coordinate system.}
\label{Fig3}
\end{figure}

Since the Y-AFM is a type of helical spin order which breaks inversion symmetry, i.e. clockwise vs counterclockwise [~\ref{Fig3}(a,b)], previous studies of triangular-lattice antiferromagnets with a helical spin order have found the magnetism induced FE $P$ \cite{kimura2006inversion,seki2007impurity,singh2009agcrs2,xiang2011general}. Thus it is reasonable to expect the similar multiferroicity in the Hf$_{2}$VC$_{2}$F$_{2}$ monolayer.

For the Y-AFM spin order, the noncollinear spin texture forms a helical plane. It is necessary to know the easy plane/axis first. Our calculation with SOC finds that the out-of-plane $c$ direction is the easy axis. Thus in the ground state, the helical plane should be perpendicular to the monolayer. The energy of $ac$ (or $bc$) plane Y-AFM is lower than that of $ab$ plane Y-AFM by $0.14$ meV/V. Our calculation also finds the rotation symmetry within the monolayer plane. Thus the helical plane can be rotated freely along the $c$-axis, as sketched in ~\ref{Fig3}(c).

The standard Berry phase calculation with SOC gives $1.98\times10^{-6}$ $\mu$C/m for the Y-AFM state, corresponding to $2700$ $\mu$C/m$^2$ in the 3D unit considering the thickness of monolayer $7.0$ {\AA}. To partition these two contributions, using the high-symmetric crystalline structure, the obtained pure electronic contribution ($P_e$) is about $1.95\times10^{-6}$ $\mu$C/m, very close to the total $P$ with ionic displacements. Therefore, here FE $P$ is almost fully ($\sim98.5\%$) originated from the bias of electronic cloud while the atomic structure is almost in the high symmetric one. Our calculation also indicates that the direction of $P$ is always perpendicular to the spin helical plane, as sketched in ~\ref{Fig3}(c). And this $P$ can be switched to $-P$, once the chirality of Y-AFM is reversed. For comparison, the higher energy $ab$-plane Y-AFM gives $2.9\times10^{-7}$ $\mu$C/m, pointing along the $c$-axis.

Although this $P$ is much smaller than those of other 2D FE's, e.g. for some functionalized 2D materials ($3\times10^{-5}-1.1\times10^{-4}$ $\mu$C/m \cite{wu2016ferroelectricity}), 2D honeycomb binary buckled compounds ($9\times10^{-7}-1.11\times10^{-5}$ $\mu$C/m \cite{di2015emergence}), and multiferroic C$_6$N$_8$H organic network ($\sim4500$ $\mu$C/m$^2$ \cite{tu2017two}), it should be noted that the origin of FE $P$ in Hf$_{2}$VC$_{2}$F$_{2}$ monolayer is conceptually different from other 2D FE's. In fact, it is common sense that the improper FE $P$'s in the type-II multiferroics are weaker than those typical values of proper FE's \cite{cheong2007multiferroics,dong2015multiferroic}. Even though, the origin of ferroelectricity in the type-II multiferroics guarantees the intrinsically strong magnetoelectric coupling, which is rare in other multiferroics.

In fact, $2700$ $\mu$C/m$^2$ is already a very significant value in type-II multiferroics, especially considering the fact that its origin is from the SOC, not exchange striction \cite{cheong2007multiferroics,dong2015multiferroic}. For reference, the $P$ in polycrystal Ba$_3$MnNb$_2$O$_9$ only reaches $3.45$ $\mu$C/m$^2$ \cite{lee2014magnetic}, and $\sim600$ $\mu$C/m$^2$ in TbMnO$_3$  \cite{Kimura:Nat}. The relatively large $P$ is probably due to $5d$ Hf ions, which own larger SOC than $3d$ elements. Although Hf's orbitals do not contribute to magnetism directly, the hybridization between orbitals always exists around the Fermi level, which may enhance the effective SOC. Thus in principle, the macroscopic polarization should be detectable, at least in its corresponding bulk form. In addition, the second-harmonic-generation (SHG) based on nonlinear optical process can also be employed to detect the polarization and its domain without electrodes, as done for TbMnO$_3$ \cite{matsubara2015magnetoelectric}. Among type-II multiferroics, some polarizations are generated by noncollinear spin order via spin-orbit coupling (SOC), as in our Hf$_{2}$VC$_{2}$F$_{2}$. Since SOC is usually weak especially for 3$d$ electrons, the polarizations in this category are usually much smaller than those in conventional ferroelectrics \cite{dong2015multiferroic}. Furthermore, here the high ratio of $P_e$/$P$ is also advantage for ultra-fast switching. Although it's common sense that type-II multiferroics own high $P_e$/$P$ (e.g. $\sim25\%$ in TbMnO$_3$ \cite{walker2011femtoscale}, and $\sim58\%$ in HoMnO$_3$ \cite{Picozzi:Prl},  which were estimated using the same method used here) than proper FE materials, the $P_e$/$P$=$98.5\%$ is indeed very high and rather rare.

The origin of ferroelectricity driven by helical spin order is also nontrivial. Although the spin-current model (or the Dzyaloshinskii-Moriya interaction) can explain the origin of ferroelectricity in cycloid spiral magnets \cite{Katsura:Prl,Sergienko:Prb}, its equation $\textbf{e}_{ij}\times(\textbf{S}_i\times\textbf{S}_j)$ gives zero net $P$ for Y-AFM in each triangular unit. Instead, the generalized spin-current model proposed by Xiang \textit{et al.} \cite{Xiang:Prl11} can phenomenologically explain the origin of $\textbf{P}$:
\begin{equation}
\textbf{P}=\textbf{M}\cdot\sum_{<ij>}(\textbf{S}_i\times\textbf{S}_j),
\label{eq1}
\end{equation}
where the summation is over all NN bonds; $\textbf{S}$ denotes a (normalized) spin vector; $\textbf{M}$ is a $3\times3$ matrix which can be determined via DFT:
\begin{equation}
\textbf{M}=-\begin{bmatrix}
9.571 &0  &0 \\
0 & 9.571  &0 \\
0 &0  &1.401
\end{bmatrix}\times10^{-3} {\rm e{\AA}}.
\end{equation}

\begin{figure}
\centering
\includegraphics[width=0.6\textwidth]{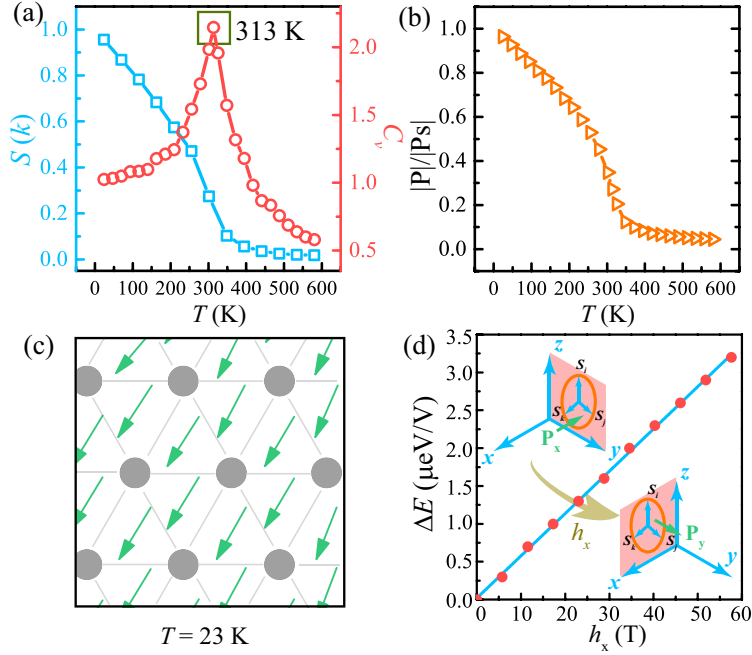}
\caption{MC results. (a) The spin structure factor ($S$($\textbf{k})=\sum_{ij}\left \langle  S_i\cdot S_{j}\right \rangle e^{\textbf{k}\cdot(r_i-r_j)}$) for Y-AFM and specific heat ($C_{\rm v}$) as a function of temperature ($T$). (b) The FE $|\textbf{P}|$ as a function of $T$, calculated using Eq.~~\ref{eq1} and normalized to its saturated value $|\textbf{P}_S|$. (c) A typical MC snapshot of $\textbf{P}_i$ (arrows) in a small region. Dots: V ions. (d) Energy difference between the $yz$ plane and $xz$ plane Y-AFM, under a magnetic field $h_x$ along the $x$ axis.}
\label{Fig4}
\end{figure}

\textit{MC simulation.}In above DFT calculations, only five magnetic candidates were considered, which could not exclude other possible exotic orders. Thus the unbiased Monte Carlo (MC) simulation is performed to verify the ground state and estimate the transition temperature \cite{Supp}. The Heisenberg spin model is adopted:
\begin{equation}
H=-J_1\sum_{<ij>}\textbf{S}_i\cdot\textbf{S}_j-J_2\sum_{[kl]}\textbf{S}_k\cdot\textbf{S}_l-A\sum_i(S_i^z)^2,
\end{equation}
where $J_1$ ($J_2$) is the exchange interaction between NN (NNN) spin pairs; $A$ is the coefficient for magnetocrystalline anisotropy and $S^z$ is the component of spin along the magnetic easy axis. Using the normalized $|\textbf{S}|=1$, these coefficients can be extracted from DFT calculations by comparing the energies of magnetic candidates \cite{Supp}: $J_1=-48.1$ meV, $J_2=6.7$ meV, and $A=0.14$ meV, respectively. As expected, the NN exchange is strongly AFM, while the NNN is much weaker. The dominant $J_1$ leads to the Y-AFM, as confirmed using MC simulation [~\ref{Fig4}(a)]. Interestingly, the estimated N\'eel temperature $T_{\rm N}$ reaches $313$ K, a remarkable high $T_{\rm N}$ above room temperature. The FE $|\textbf{P}|$ just appears below $T_{\rm N}$ [~\ref{Fig4}(b)], a character of type-II multiferroicity. A typical MC snapshot of local $\textbf{P}_i$'s (of V triangular units) at low temperature is shown in \ref{Fig4}(c). 

For most 3D type-II multiferroics, the magnetism and ferroelectricity only appear far below room temperature \cite{dong2015multiferroic}. High-temperature type-II multiferroicity is a highly desired property for applications, which is a bottleneck for this category of materials. Till now, in various type-II multiferroics, only a few hexagonal ferrites with very complex crystalline/magnetic structures show magnetoelectricity above room temperature \cite{Kimura:Arcp}. Hf$_{2}$VC$_{2}$F$_{2}$ is another room-temperature type-II multiferroic system, with a much simpler crystalline/magnetic structure.

Physically, its high $T_{\rm N}$ is due to the ideal half-filled $t_{\rm 2g}$ orbitals ($3d^3$), which prefers a strong superexchange according to the Goodenough-Kanamori rule \cite{goodenough1958interpretation,kanamori1959superexchange}. The similar case is for various ferrites with Fe$^{3+}$ ($3d^5$) ions, which usually own magnetic orders above room temperature. 

As a type-II multiferroic, the induced $\textbf{P}$ can be fully controlled by magnetic fields via the helical plane rotation \cite{Supp}. As shown in ~\ref{Fig4}(d), under an in-plane magnetic field, the energies of $yz$-plane and $xz$-plane Y-AFM (after slight distortions driven by magnetic field) are no longer degenerated. Thus, the helical plane of Y-AFM and its associated $\textbf{P}$ should rotate accompanying the field. Since there's no intrinsic energy barrier for such a helical plane rotation, this magnetoelectric response should work under small fields.

Last, the MoSe$_2$ substrate is considered to test the possible substrate effect \cite{Supp}. With proximate in-plane lattice constants, the optimized distance between Hf$_2$VC$_2$F$_2$ monolayer and MoSe$_2$ substrate is $\sim3.5$ \AA{}, indicating the vdW interaction. No charge transfer occurs between Hf$_2$VC$_2$F$_2$ and substrate. The Y-AFM remains the ground state, and the whole system remains insulating. Therefore, the substrate will not change the conclusion of multiferroicity for Hf$_2$VC$_2$F$_2$ monolayer.

\textit{Conclusion.} The noncollinear $120^\circ$ Y-type antiferromagnetic order is predicted to be the ground state in MXene Hf$_{2}$VC$_{2}$F$_{2}$ monolayer, and the estimated N\'eel point can be above room temperature. More importantly, the inversion symmetry is broken by this particular Y-type antiferromagnetic order, resulting in the improper magnetism-driven ferroelectric polarization. Thus Hf$_{2}$VC$_{2}$F$_{2}$ monolayer is a room-temperature type-II multiferroics, which has intrinsically strong magnetoelectric coupling. The crossover between 2D materials and magnetic ferroelectrics will be a very interesting topic, both fundamentally and to benefit nanoscale devices.

\begin{acknowledgement}
Work was supported by National Natural Science Foundation of China (Grant No. 11674055), Fundamental Research Funds for the Central Universities, Jiangsu Innovation Projects for Graduate Student (Grant No. KYLX16\underline{ }0116), and National Supercomputer Center in Guangzhou (NSCC-GZ).
\end{acknowledgement}
\bibliography{ref}
\end{document}